\documentclass[aps,amsmath,amssymb,superscriptaddress,nofootinbib]{revtex4-2}
\usepackage{graphicx}
\usepackage{hyperref}
\usepackage{siunitx}
\usepackage{lineno}
\usepackage{graphicx}
\usepackage{amsmath}
\usepackage{color}
\usepackage{bm}
\usepackage{amssymb}
\usepackage{mathrsfs}
\usepackage{amsbsy}
\usepackage{subcaption}
\newcommand{\meqref}[1]{Eq.(\ref{#1})}
\newcommand{\mpref}[1]{Fig.\ref{#1}}
\newcommand{\mtref}[1]{Tab.\ref{#1}}
\newcommand{\be}{\begin{equation}}
\newcommand{\ee}{\end{equation}}

\begin{document}
\title{Circular Motion of Charged Particles near Charged Black Hole}
\author{Yu-Qi Lei}
\affiliation{Department of Physics, Shanghai University, Shanghai 200444, China}
\author{Xian-Hui Ge}
\email{gexh@shu.edu.cn}
\affiliation{Department of Physics, Shanghai University, Shanghai 200444, China}
\affiliation{
Center for Gravitation and Cosmology, College of Physical Science and Technology,
Yangzhou University, Yangzhou 225009, China}

\begin{abstract}
We study the circular motion of charged particles near a charged black hole. The general form of the Lyapunov exponent of charged particles is obtained by using the Jacobian matrix. The results show that the chaos bound can be saturated by the circular motion of charged particles on the horizon. By further expanding the Lyapunov exponent near the horizon and investigating Reissner-Nordstr\"om(RN) black holes with different $M/Q$, we find that in contrast to the static equilibrium, the circular motion of charged particles can have a larger Lyapunov exponent due to the existence of angular momentum. For the RN black holes which have the mass-charge ratio $1<M/Q <1.1547$, the chaos bound is locally violated by the null and the time-like circular motion with large angular momentum. As an illustration of the universality of our results, we study the charged particles' circular motion near the Reissner-Nordstr\"om Anti-de Sitter (RN-AdS) black hole and find that the null and the time-like circular motion with large angular momentum can exceed the chaos bound under the background of RN-AdS black holes with the mass-charge ratio in the range $1.23132<M/Q<1.75225$.
\end{abstract}

\maketitle

\section{Introduction}
Recently, much attention has been paid to the chaotic phenomenon of particle motion near black holes, such as the study of distinguishing the chaotic orbits of particles from periodic orbits \cite{Suzuki_199,letelier1997chaos,aless1999chaos,dalui2019induction,Kao:2004qs,Chen:2016tmr,Wang:2016wcj,Ma:2014aha} and the stability analysis of particle motion near black holes \cite{Cardoso:2008bp,Pradhan:2012rkk,Pradhan:2013bli}. These studies can help us to recognize the aspects of chaos in the black hole theory, and what the properties of black hole chaos need to be further explored. 

A general upper bound of chaos in quantum systems has been proposed by Maldacena, Shenker and Standford from the quantum field theory, which indicates that the Lyapunov exponent $\lambda$, describing the strength of chaos, has a temperature-dependent upper bound \cite{Maldacena:2015waa}
\be
\lambda \leq \frac{2 \pi T}{\hbar}.\label{cb0}
\ee
This temperature-dependent ansatz was examined by shock wave gedanken-experiments \cite{Shenker:2013pqa,Shenker:2013yza} and the AdS/CFT correspondence \cite{Maldacena:1997re}. With the black hole thermodynamic relationship $\kappa =2\pi T$ and setting $\hbar=1$, the chaos bound \meqref{cb0} has an equivalent form at the horizon
\be
\lambda \leq \kappa,\label{cb}
\ee
where $\kappa$ is the surface gravity of black holes. This bound may build a strong connection between quantum chaos and black hole physics. More related calculations have shown the universality of chaos bound in particle motion near black holes, such as the momentum growth of particles' radial falling into black holes \cite{Susskind:2018tei,Brown:2018kvn,Ageev:2018msv,Zhou:2018rac}, the chaos in the null orbits near the horizon \cite{Dalui:2018qqv,Bera:2021sce} and the bound of Lyapunov exponent in Kerr-Newman black hole \cite{Kan:2021blg}.

The chaos bound \meqref{cb} can not only be studied by shock waves of black holes \cite{Jahnke:2019gxr,Poojary:2018esz,Liu:2020yaf}, but also the particle motion near black holes. For example in \cite{Hashimoto:2016dfz}, the chaos bound was studied by particle motion near the black hole horizon, in which the static equilibrium of particles with external forces near the horizon has been explored to examine the chaos bound. As shown, the Lyapunov exponent of static equilibrium can saturate the chaos bound at the horizon. Meanwhile, the study on the near-horizon behavior of the charged particles' static equilibrium shows that the chaos bound may be violated \cite{Zhao:2018wkl}, and in the systems where chaos bound can be violated, the chaos is strengthened \cite{Lei:2020clg}.

It would be of interesting to investigate the circular motion of charged particles near charged black holes in which the electromagnetic force and the circular motion can maintain equilibrium. The advantage of this study is that it can be done simply for Reissner-Nordstr\"om (RN) black holes. We study the circular motion of charged particles in the area near the charged black holes i.e. the region between the outer horizon radius $r_+$ and the photon sphere radius $r_{ps}$, where the photon sphere radius refers to the innermost circular orbital radius of the neutral particle. We calculate the Lyapunov exponents of charged particles' circular motion near the Reissner-Nordstr\"om (RN) black holes with different mass-charge ratio $M/Q$, and the results show that for the near-extremal RN black holes with $M/Q <1.1547$, the chaos bound can be violated by the null circular motion and the time-like circular motion with large momentum, in the region $r_+<r<r_{ps}$. In order to check the universality of our results, we also consider the circular motion of charged particles in the background of Reissner-Nordstr\"om Anti-de Sitter (RN-AdS) black holes. In the range $1.23132<M<1.75225$ and $r_+<r<r_{ps}$, the chaos bound can also be violated. The results show that as the black hole approaches zero temperature ($\kappa \rightarrow 0$), the Lyapunov exponent of charged particles' circular motion near horizon can in fact violate the chaos bound.

From the calculations of RN black hole and RN-AdS black hole, we can see that the time-like circular motion of charged particles becomes unstable as the angular momentum increases. Here, we focus on examining the relationship between the Lyapunov exponent of circular motion and the chaos bound, instead of analyzing the Poincar\'e section. The discussion about the Poincar\'e section is carried out in Refs.\cite{Hashimoto:2016dfz,Dalui:2018qqv,Lei:2020clg,Bera:2021sce}.

The organization of this paper is as follows. In section \ref{general form}, we give a general form of the calculation about the charged particles' circular motion near charged black holes, including the equation of motion and the circular motion condition. In section \ref{Ly and chaos bound}, the method of calculating the Lyapunov exponent through the Jacobian matrix is shown, and the relationship between the circular motion and the chaos bound is discussed. Meanwhile, we propose some parameters to predict whether the chaos bound is violated in near-horizon expansion. In section \ref{RN}, we discuss the violation of chaos bound from the charged particles' circular motion in RN black holes with different $M/Q$. The circular motion near Reissner-Nordstr\"om Anti-de Sitter (RN-AdS) black hole is also studied in section \ref{RNAdS}. We conclude this paper in section \ref{conclusion}.

\section{The charged particles' circular motion near a charged black hole}\label{general form}
In this section, we investigate the circular motion of charged particle in general charged black hole background. Considered the Einstein-Maxwell gravity, its Lagrangian can be written as
\be
\mathcal{L}=\sqrt{-g} \left(R-\frac{1}{4}F^2+2\Lambda_0 \right),
\ee
which allows a 4-dimensional static spherically symmetric charged black hole solution with the metric
\be
ds^2=-f(r)dt^2+\frac{dr^2}{f(r)}+r^2(d\theta^2+\sin^2\theta d\phi^2),
\ee
where $f(r)$ is the metric function. There are two horizons ($r_-$ and $r_+$) defined by $f(r_-)=f(r_+)=0$ respectively. With $r_+>r_-$, $r_+$ is the outer horizon of this charged black hole. At the outer horizon, the surface gravity $\kappa$ can be obtained by $\kappa =\left.\frac{1}{2}f^{'} \right|_{r=r_+}$, where the prime denotes a derivative with respect to $r$. A Maxwell field $A=A_\mu dx^\mu$ is assumed to exist, where $A_\mu=(A_t(r),\ 0,\ 0,\ 0)$.

The Lagrangian of a test particle with charge $q$ moving on the equatorial plane($\dot{\theta}=0, \theta=\frac{\pi}{2}$) of charged black hole can be written as
\be
\mathcal{L}=\frac{1}{2}\left(-f\dot{t}^2+\frac{\dot{r}^2}{f}+r^2\dot{\phi}^2 \right)-A_t\dot{t},
\ee
where the dot denotes a derivative with respect to the proper time $\tau$. Then we can define the generalized momenta by $\pi_\mu=\frac{\partial \mathcal{L}}{\partial \dot{x}^\mu}$. These generalized momenta are
\be
\begin{aligned}
\pi_t=&-f\dot{t}-q A_t=-E=Const,
\\
\pi_r=&\frac{\dot{r}}{f},
\\
\pi_\phi=&r^2\dot{\phi}=L=Const,
\end{aligned}
\ee
where $E$ and $L(\pi_\phi)$ are the total energy and the angular momentum of this test particle, respectively.

The Hamiltonian of the test particle can be given by $H=\pi_\mu \dot{x}^\mu-\mathcal{L}$, which is
\be
H=\frac{1}{2}\left(-\frac{(\pi_t+qA_t)^2}{f}+f\pi_r^2+\frac{\pi_\phi^2}{r^2} \right).
\ee
The Hamiltonian equations of particle motion are obtained in a standard way
\be
\dot{x}^{\mu}=\frac{\partial H}{\partial \pi_\mu}, \qquad \dot{\pi}_{\mu}=-\frac{\partial H}{\partial x^{\mu}}.
\ee
These equations are
\be
\begin{aligned}
\dot{t}=&-\frac{\pi_t+qA_t}{f},
\quad
\dot{r}=\pi_rf,
\quad
\dot{\phi}=\frac{\pi_\phi}{r^2},
\\
\dot{\pi}_t=&0,
\quad
\dot{\pi}_r=\frac{\pi_\phi^2}{r^3}-\frac{\pi_r^2}{2f^{'}}+\frac{(\pi_t+qA_t)(2qfA_t^{'}-f^{'}(\pi_t+qA_t))}{2f^2},
\quad
\dot{\pi}_\phi=0.
\end{aligned}\label{EOM}
\ee

Next, we consider the constraints for charged particles' circular motion on the black hole equatorial plane. For the circular motion, it obeys the radial equilibrium condition $\frac{dr}{dt}=\frac{d\pi_r}{dt}=0$\footnote{Since the Lyapunov exponent is related to the choice of the time coordinates and the chaos bound is defined with the coordinate time $t$, so we discuss the radial equilibrium of particles at coordinate time $t$. In addition, $\frac{dr}{dt}=\frac{\dot{r}}{\dot{t}}$, $\frac{d\pi_r}{dt}=\frac{\dot{\pi_r}}{\dot{t}}$.}. With the radial equilibrium condition and \meqref{EOM}, we then obtain
\be
\pi_\phi=\frac{\sqrt{r^3(\pi_t+qA_t)((\pi_t+qA_t)f^{'}-2qf^{'}A_t^{'})}}{\sqrt{2}f}.\label{pphi0}
\ee
Another constraint is from the normalization of the four-velocity ($\dot{x}^\mu$)
\be
g_{\mu \nu}\dot{x}^\mu \dot{x}^\nu=\eta.\label{ssdgy}
\ee
The constant $\eta$ depends on whether it is for massive particles or massless particles. For the massive particles, there are time-like orbits with $\eta=-1$, and the massless particles move along null orbits with $\eta =0$.

For the massive particles, they move along time-like orbits with $\eta=-1$. For the circular motion with $\frac{dr}{dt}=\frac{d\pi_r}{dt}=0$, we can obtain
\be
\pi_t=-qA_t+\frac{\sqrt{f(\pi_\phi^2+r^2)}}{r}.\label{pt0tl}
\ee
With \meqref{pphi0} and \meqref{pt0tl}, the charge $q$ of massive particle's circular motion should satisfies
\be
q=\frac{-2\pi_\phi^2f+rf^{'}(\pi_\phi^2+r^2)}{2r^2A_t^{'}\sqrt{f(\pi_\phi^2+r^2)}}.\label{q0tl}
\ee

For the massless particles, their null orbits satisfy $\eta=0$. So the null circular orbits ($\frac{dr}{dt}=\frac{d\pi_r}{dt}=0$) should satisfy
\be
\pi_t=-qA_t+\frac{\pi_\phi \sqrt{f}}{r}.\label{pt0n}
\ee
From \meqref{pphi0} and \meqref{pt0n}, we can see the charge of particles for null circular orbits should obey
\be
q=\frac{\pi_\phi(rf^{'}-2f)}{2r^2A_t^{'}\sqrt{f}}.\label{q0n}
\ee

In the above, the circular motion conditions of charged particles (massive and massless) have been discussed. Then we will show how to use the Lyapunov exponent to analyze the stability of circular motion and discuss its relation to the upper chaos bound $\lambda \leq \kappa$.

\section{Lyapunov exponent and chaos bound}\label{Ly and chaos bound}
In this section, we use the Jacobian matrix method \cite{Cardoso:2008bp,Pradhan:2012rkk,Pradhan:2013bli} to calculate the Lyapunov exponent of the charged particles' circular motion near charged black holes. Then, the relationship between the Lyapunov exponent and the chaos bound $\lambda \leq \kappa$ will be explored.

\subsection{Lyapunov exponent}
\qquad From the charged particles' equation of motion \meqref{EOM}, we can obtain the radial equation at coordinate time $t$
\be
\begin{aligned}
\frac{dr}{dt}=&\frac{\dot{r}}{\dot{t}}=F_1(r,\pi_r)=-\frac{\pi_rf^2}{\pi_t+qA_t},
\\
\frac{d\pi_r}{dt}=&\frac{\dot{\pi}_r}{\dot{t}}=F_2(r,\pi_r)=-qA_t^{'}+\frac{(\pi_t+qA_t)f^{'}}{2f}+\frac{f(-2\pi_\phi^2+\pi_r^2r^3f^{'})}{2r^3(\pi_t+qA_t)}.
\end{aligned}
\ee
Taking $(r,\pi_r)$ as the phase space variables, we can obtain the Jacobian matrix $K_{ij}$
\be
K_{11}=\frac{\partial F_1}{\partial r},
\quad
K_{12}=\frac{\partial F_1}{\partial \pi_r},
\quad
K_{21}=\frac{\partial F_2}{\partial r},
\quad
K_{22}=\frac{\partial F_2}{\partial \pi_r}.\label{JM0}
\ee
When the circular motion of particles outside black holes is considered, we can obtain $\pi_r=0$, then the Jacobian matrix $K_{ij}$ can be reduced to
\be
K_{ij}=
\left(
\begin{matrix}
0&K_{12}
\\
K_{21}&0
\end{matrix}
\right),
\ee
the eigenvalues of the Jacobian matrix $K_{ij}$ lead to the Lyapunov exponent $\lambda$ 
\be
\lambda=\sqrt{K_{12} K_{21}}.
\ee
The stability of the circular motion can be specified by $\lambda^2$. It can be shown as
\begin{equation*}
\left\{
\begin{aligned}
\lambda^2>0&: {\rm \quad the\ circular\ motion\ is\ unstable;}
\\
\lambda^2=0&: {\rm \quad the\ circular\ motion\ is\ marginal;}
\\
\lambda^2<0&: {\rm \quad the\ circular\ motion\ is\ stable.}
\end{aligned}
\right.
\end{equation*}
For the unstable case, once circular motion is perturbated, the perturbation will increase exponentially, which indicates the presence of chaos. Meanwhile, the larger the Lyapunov exponent $\lambda$, the stronger the chaos. Our main purpose is to explore the relationship between the Lyapunov exponent of charged particles' circular motion and the chaos bound $\lambda \leq \kappa$. Let us first calculate the Lyapunov exponent of the circular motion of charged particles.

For the massive particle, from $\frac{dr}{dt}=\frac{d\pi_r}{dt}=0$, \meqref{pt0tl} and \meqref{q0tl}, the components of the Jacobian matrix $K_{ij}$ can be simplified as

\be
\begin{aligned}
K_{11}=&0,
\\
K_{12}=&-\frac{r f^{\frac{3}{2}}}{\sqrt{\pi_\phi^2+r^2}},
\\
K_{21}=&\frac{rA_t^{'}f^{''}(\pi_\phi^2+r^2)-(2\pi_\phi^2A_t^{'}+rA_t^{''}(\pi_\phi^2+r^2))f^{'}}{2r^2A_t^{'}\sqrt{f(\pi_\phi^2+r^2)}}
\\
+&\frac{\pi_\phi^2f^2(A_t^{'}(2\pi_\phi^2+3r^2)+rA_t^{''}(\pi_\phi^2+r^2))}{r^3A_t^{'}(f(\pi_\phi^2+r^2))^{\frac{3}{2}}}-\frac{\sqrt{\pi_\phi^2+r^2}f^{'2}}{4rf^{\frac{3}{2}}},
\\
K_{22}=&0.
\end{aligned}\label{ktl}
\ee

Then the eigenvalues of $K_{ij}$ can give the Lyapunov exponent $\lambda$. Therefore, we can obtain the Lyapanov exponent of the time-like circular motion from \meqref{ktl}
\be
\begin{aligned}
\lambda_{tc}^2=&\frac{1}{4}f^{'2}-\frac{\pi_\phi^2f^2\left(A_t^{'}(2\pi_\phi^2+3r^2)+rA_t^{''}(\pi_\phi^2+r^2) \right)}{r^2A_t^{'}\left(\pi_\phi^2+r^2 \right)^2}
\\
&+\frac{f}{4}\left(f^{'}\left(\frac{4\pi_\phi^2}{\pi_\phi^2+r^3}+\frac{2A_t^{''}}{A_t^{'}}\right)-2f^{''} \right)
\end{aligned}\label{ltc}
\ee
This formula embodies the relationship between the Lyapunov exponent and the angular momentum $\pi_\phi$.

For the massless particle satisfying $\frac{dr}{dt}=\frac{d\pi_r}{dt}=0$, \meqref{pt0n} and \meqref{q0n}, we can simplify the Jacobian matrix $K_{ij}$ as
\be
\begin{aligned}
K_{11}=&0,
\\
K_{12}=&-\frac{rf^{\frac{3}{2}}}{\pi_\phi},
\\
K_{21}=&\frac{\pi_\phi(-r^2A_t^{'}f^{'2}+4f^2(2A_t^{'}+rA_t^{''}))}{4r^3A_t^{'}f^{\frac{3}{2}}}
\\
+&\frac{f\pi_\phi(-f^{'}(2A_t^{'}+rA_t^{''})+rA_t^{'}f^{''})}{2r^3A_t^{'}f^{\frac{3}{2}}},
\\
K_{22}=&0.
\end{aligned}\label{kn}
\ee
So the Lyapunov exponent $\lambda_{nc}$ of null circular motion should satisfies
\be
\lambda_{nc}^2=\frac{1}{4}\left(f^{'2}-\frac{4f^2(2A_t^{'}+rA_t^{''})}{r^2A_t^{'}}+f\left(f^{'}\left(\frac{4}{r}+\frac{2 A_t{''}}{A_t^{'}}\right)-2f^{''} \right) \right).\label{lnc}
\ee

It is easy to verify that when $\pi_\phi \rightarrow \infty$, $\lambda_{tc}$ in \meqref{ltc} will be reduced to $\lambda_{nc}$. Therefore, the behavior of the null circular orbits discussed here can be regarded as a limit case of the time-like circular orbits. At the outer horizon, we have $f(r_+)=0$, which leads to $\lambda_{tc}=\lambda_{nc}=\kappa$. That is to say, the chaos bound $\lambda \leq \kappa$ is saturated at the event horizon. 

\subsection{Near-horizon behavior}
\qquad The near-horizon geometry can give us a clearer explaination, and the geometry is specified by the metric function and potential function whose Taylor expansions on the outer horizon $r=r_+$ are given by
\be
\begin{aligned}
f(r)=&f_1(r-r_+)+f_2(r-r_+)^2+f_3(r-r_+)^3+\cdots 
\\
A_t(r)=&A_{t0}+A_{t1}(r-r_+)+A_{t2}(r-r_+)^2+A_{t3}(r-r_+)^3+\cdots
\end{aligned}
\ee

To further discuss whether the chaos bound can be violated by charged particles' circular motion, we can expand the Lyapunov exponent $\lambda$ near the outer event horizon $r_+$
\begin{equation*}
\lambda^2=\kappa^2+\gamma_1(r-r_+)+\gamma_{2}(r-r_+)^2+\sum^{\infty}_{N=3}\gamma_N (r-r_+)^N,	
\end{equation*}
where $\gamma_1$ and $\gamma_2$ are two coefficients. The relation between the Lyapunov exponent $\lambda$ and the surface gravity $\kappa$ can be determined by 
$$
\left\{
\begin{aligned}
\lambda<\kappa& \qquad \gamma_{1}(r-r_+)+\sum^{\infty}_{N=3}\gamma_N (r-r_+)^N<0;
\\
\lambda=\kappa& \qquad \gamma_{1}(r-r_+)+\sum^{\infty}_{N=3}\gamma_N (r-r_+)^N=0;
\\
\lambda>\kappa& \qquad \gamma_{1}(r-r_+)+\sum^{\infty}_{N=3}\gamma_N (r-r_+)^N>0.
\end{aligned}
\right.
$$
Then the expression of $\lambda_{tc}$ and $\lambda_{nc}$ can be rewritten as
\be
\begin{aligned}
\lambda_{tc}^2=&\kappa^2+\gamma_{t1}(r-r_+)+\gamma_{t2}(r-r_+)^2+\sum^{\infty}_{N=3}\gamma_{tN} (r-r_+)^N,
\\
\lambda_{nc}^2=&\kappa^2+\gamma_{n1}(r-r_+)+\gamma_{n2}(r-r_+)^2+\sum^{\infty}_{N=3}\gamma_{nN} (r-r_+)^N.
\end{aligned}\label{lye}
\ee
where we have
\be
\begin{aligned}
\gamma_{t1}=&f_1^2 \left(\frac{\pi_\phi^2}{\pi_\phi^2 r_+ +r_+^3}+\frac{A_{t2}}{A_{t1}}\right),
\\
\gamma_{t2}=&\frac{f_1((\pi_\phi^2r_++r_+^3)(3A_{t3}f_1+3A_{t2}f_2)-2A_{t2}f_1\pi_\phi^2)}{A_{t1}(\pi_\phi^2r_++r_+^3)}
\\
&-\frac{3f_1(f_3r_+(\pi_\phi^2+r_+^2)-2f_2\pi_\phi^2)}{2r_+(\pi_\phi^2+r_+^2)}-\frac{2A_{t2}^2f_1^2}{A_{t1}^2}-\frac{3f_1^2\pi_\phi^2}{r_+^2(\pi_\phi^2+r_+^2)},
\\
\gamma_{n1}=&\frac{f_1^2(A_{t1}+A_{t2}r_+)}{A_{t1}r_+},
\\
\gamma_{n2}=&-\frac{2A_{t2}^2f_1^2}{A_{t1}^2}+\frac{f_1(3A_{t3}f_1r_++3A_{t2}f_2r_+-A_{t2}f_1)}{A_{t1}r_+}
\\
&-\frac{3f_1(2f_1+r_+(f_3r_+-2f_2))}{2r_+^2}.
\end{aligned}
\ee

From \meqref{lye}, we can see that the parameters $\gamma_{t1}$, $\gamma_{t2}$, $\gamma_{n1}$ and $\gamma_{n2}$ can determine whether the circular motion of charged particles in the near-horizon geometry can violate the chaos bound. Note that the near-horizon expansion is not the unique condition for judging the violation of the chaos bound. Once we obtain the analytical expression of the Lyapunov exponent $\lambda$, we can compare the value of $\lambda$ and $\kappa$ in the whole region outside the horizon.

\section{Reissner-Nordstr\"om (RN) black hole}\label{RN}
After discussing the general form of the charged particles' circular motion outside the charged black holes, we specifically study the Lyapunov exponent of charged particles in the 4-dimensional RN black hole background. The metric function $f(r)$ and electric potential function $A_t(r)$ are \cite{Zhao:2018wkl}
\be
f(r)=1-\frac{2M}{r}+\frac{Q^2}{r^2},\qquad A_t(r)=\frac{2Q}{r},
\ee
where $M$ is the black hole mass and $Q$ is the charge of the black hole. If $M=Q$, this is an extremal RN black hole with the black hole temperature $T=0$. For the non-extremal RN black hole ($M>Q$), there are two horizons at $r=r_+$ and $r=r_-$ (where $r_+>r_-$). Then the parameters ($M$, $Q$) and the surface gravity on the outer horizon $r_+$ can be expressed as
\be
M=\frac{r_+ +r_-}{2},\quad Q=\sqrt{r_+r_-}, \quad \kappa=\frac{r_+-r_-}{2r_+^2}.\label{rnmq}
\ee
When $M<Q$, there will be a naked singularity. To avoid this situation, we require $M/Q>1$.

\subsection{Near-horizon analysis}
Then the parameters ($\gamma_{t1}$, $\gamma_{t2}$, $\gamma_{n1}$ and $\gamma_{n2}$) in near-horizon expansion can be rewritten as
\be
\begin{aligned}
\gamma_{t1}=&-\frac{4(Q^2-Mr_+)^2}{r_+^5(\pi_\phi^2+r_+^2)},
\\
\gamma_{t2}=&\frac{10(Q^2-Mr_+)^2}{r_+^4(\pi_\phi^2+r_+^2)^2}-\frac{6\pi_\phi^4(2Q^4-3MQ^2r_++M
^2r_+^2)}{r_+^8(\pi_\phi^2+r_+^2)^2}
\\
&-\frac{2\pi_\phi^2(7Q^4-11MQ^2r_++4M^2r_+^2)}{r_+^6(\pi_\phi^2+r_+^2)^2},
\\
\gamma_{n1}=&0,
\\
\gamma_{n2}=&-\frac{6(Mr_+-2Q^2)(Mr_+-Q^2)}{r_+^8}.
\end{aligned}\label{gamrn}
\ee

For the null circular motion near RN black holes, $\gamma_{n1}=0$, so the chaos bound is violated when $\gamma_{n2}>0$. From $\gamma_{n2}$ in \meqref{gamrn}, $(Mr_+-2Q^2)(Mr_+-Q^2)<0$ is an obvious condition that the chaos bound can be violated. With the radius of outer horizon $r_+=M+\sqrt{M^2-Q^2}$, the condition can be reduced to $M/Q<\frac{2}{\sqrt{3}}=1.1547$\footnote{This is an approximation because it is obtained from the second-order near-horizon expansion.}. Then we will examine the relationship between the Lyapunov exponent of charged particles and the chaos bound based on the above discussion near RN black holes with different $M/Q$. To reflect the effect of the condition $M/Q<1.1547$, we take $M/Q=1.004,\ 1.1,\ 1.2$ in the subsequent calculations where $M/Q=1.004,\ 1.1$ meets the condition that the chaos bound can be violated. In \mtref{gn}, we show the values of $\gamma_{n1}$ and $\gamma_{n2}$ of charged particles null circular motion for different RN black holes (where $Q=1$ and $M/Q$ can be $1.004$, $1.1$ and $1.2$). For $\gamma_{n2}$, it is positive in the case of $M/Q=1.004$ and $M/Q=1.1$ and implies the violation of chaos bound $\lambda \leq \kappa$. For RN black hole with $M/Q=1.2$, $\gamma_{n2}$ is negative, which shows no violation of chaos bound.  

\begin{table}[htb]
    \centering
    \begin{tabular}{ ccc } 
        \hline
        RN black holes&$\gamma_{n1}$&$\gamma_{n2}$\\
         \hline
         M/Q=1.004 &0  &0.25916 \\ 
         \hline
         M/Q=1.1 &0  &0.03524\\
         \hline
         M/Q=1.2 &0 &-0.01204 \\ 
         \hline
    \end{tabular}
    \captionsetup{justification=raggedright}
    \caption{The parameters ($\gamma_{t1}$ and $\gamma_{t2}$) of null circular motion near different RN black holes where $Q=1$.}\label{gn}
\end{table}

\begin{figure}[htbp]
    \begin{subfigure}{3in}
      \centering
        \includegraphics[scale=0.38]{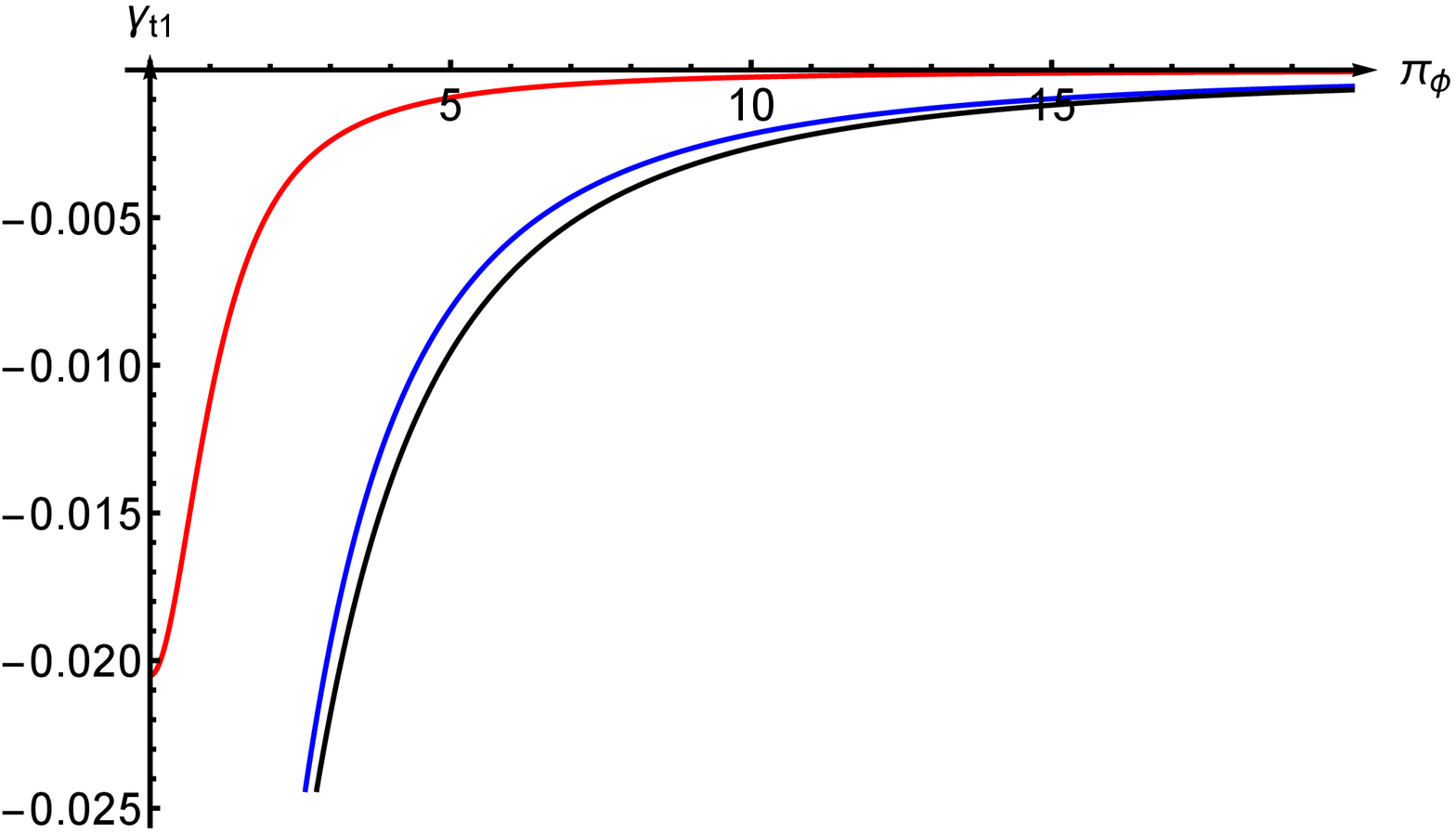}
        \caption{$\gamma_{t1}$ as a function of the angular momentum $\pi_\phi$.}\label{fig1a}
    \end{subfigure}
    \qquad 
    \begin{subfigure}{3in}
      \centering
        \includegraphics[scale=0.36]{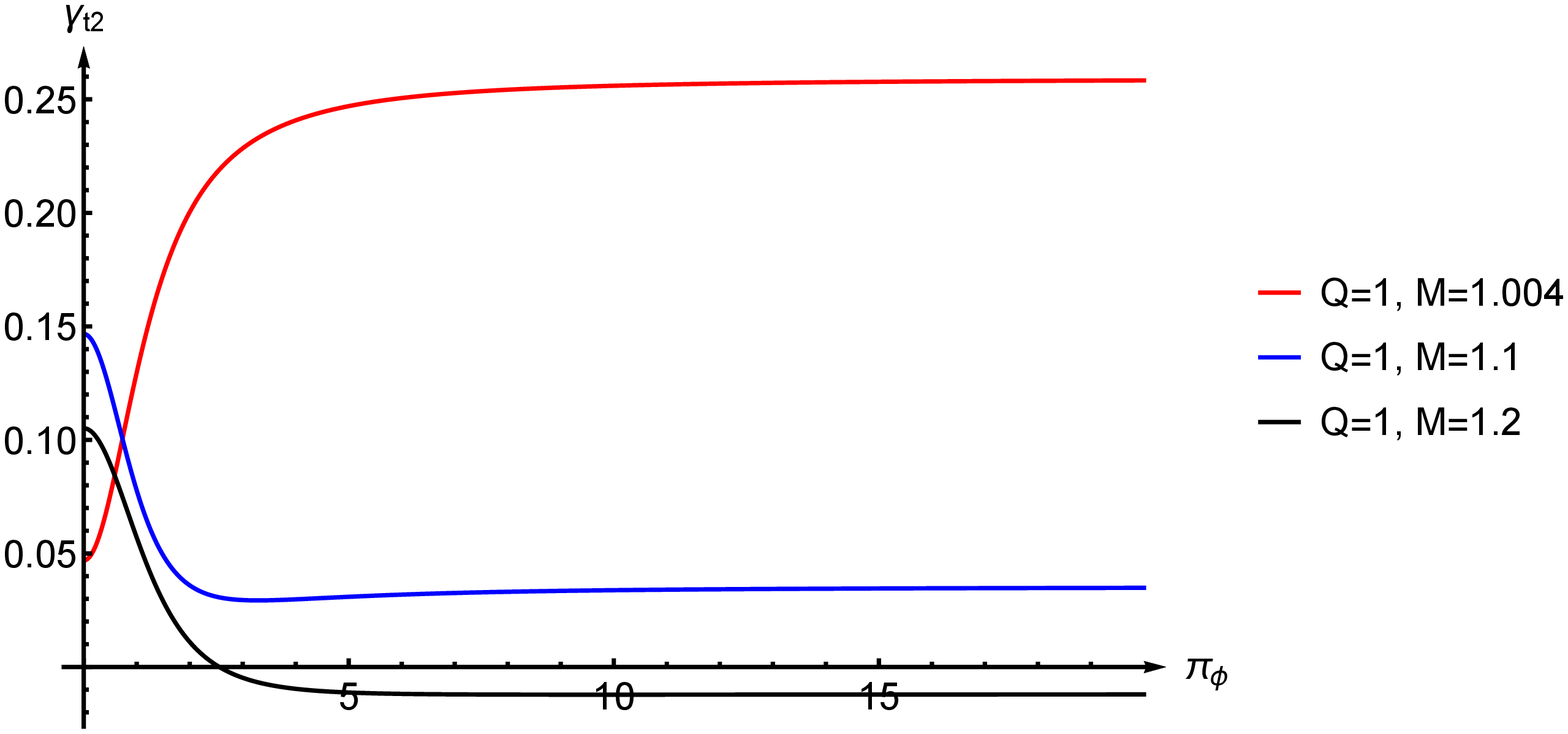}
        \caption{$\gamma_{t2}$ as a function of the angular momentum $\pi_\phi$.}\label{fig1b}
    \end{subfigure}
    \captionsetup{justification=raggedright}
    \caption{The parameters ($\gamma_{t1}$, $\gamma_{t2}$) as a function of the angular momentum $\pi_\phi$ for different mass-charge ratio $M/Q$ of RN black holes.}\label{fig1}   
\end{figure}

For the time-like circular motion near RN black holes, it is known from \meqref{gamrn} that when their angular momentum $\pi_\phi$ tend to infinity, $\gamma_{t1}$ and $\gamma_{t2}$ have the asymptotic behavior
\be 
\begin{aligned}
\lim_{\pi_\phi \rightarrow \infty}\gamma_{t1}=&\gamma_{n1}=0,
\\
\lim_{\pi_\phi \rightarrow \infty}\gamma_{t2}=&\gamma_{n2}.
\end{aligned}
\ee 
When the angular momentum of the time-like circular motion goes to infinity, its near-horizon expansion parameters $\gamma_{1}$ and $\gamma_{t2}$ tend to those of the null circular motion $\gamma_{n1}$ and $\gamma_{n2}$, which is consistent with our consideration of the null circular motion as a limiting case of the time-like circular motion. Thus we can take the result in the null circular motion as the final case of the growing angular momentum of the time-like circular motion. To better understand the Lyapunov exponential behavior in time-like circular motion, we discuss $\gamma_{t1}$ and $\gamma_{t2}$ in more detail. In \mpref{fig1} we show the relationship between ($\gamma_{t1}$, $\gamma_{t2}$) and the angular momentum of the the time-like circular motion, respectively. In \mpref{fig1a}, it is shown that $\gamma_{t1}$ is negative and tends to 0 with the growth of angular momentum. In addition, the closer $M/Q$ is to 1 ($M/Q=1$ corresponds to extreme RN black holes), the more rapidly $\gamma_{t1}$ tends to 0\footnote{More cases are discussed in the appendix.}. \mpref{fig1b} shows that $\gamma_{t2}$ changes rapidly with increasing angular momentum for small angular momentum (approximately between 0 and 3) and then converges to a certain constant value, which is positive for the case $M/Q=1.004,\ 1.1$ and negative for the case $M/Q=1.2$.

To confirm whether the time-like circular motion of charged particles can exceed the chaos bound, we need to discuss the effect of the expansion term $\gamma_{1}(r-r_+)+\gamma_{2}(r-r_+)^2$. Since the behavior of $\gamma_{t1}$ converging to zero, we define $\Delta=\frac{\gamma_{2}(r-r_+)^2}{\gamma_{1}(r-r_+)}$ to determine whether the total contribution of the expansion term is dominated by $\gamma_{t1}$ or $\gamma_{t2}$. When $\Delta>1$, the expansion term is determined by $\gamma_{t2}$, otherwise it is determined by $\gamma_{t1}$. We can consider $r=r_++\delta$ to obtain $\Delta=\frac{\gamma_{t2}\delta}{\gamma_{t1}}$, where $\delta$ is a nonzero small quantity satisfying the near-horizon expansion. Since $\gamma_{t1}$ converges to 0 as the angular momentum increases, when the angular momentum of the time-like circular motion is large enough, we can obtain $\delta \gg \gamma_{t1}$ which will lead to $\Delta>1$. At this point, if $\gamma_{t2}$ is greater than 0, the chaos bound is locally violated by time-like circular motion. Therefore, we deduce that the time-like circular motion with large angular momentum near the RN black holes ($M/Q=1.004,\ 1.1$) can violate the chaos bound. Meanwhile, in examples approaching extremal black holes, $\gamma_{t1}$ converges to 0 more rapidly, which can lead to time-like circular motion more easily exceeding the chaos bound. 

\subsection{Beyond the near-horizon region}

By calculating the parameters ($\gamma_{t1}$, $\gamma_{t2}$, $\gamma_{n1}$ and $\gamma_{n2}$), we predict that the circular motion of charged particles will violate the chaos bound. In order to support the previous results, we extend our discussions to the cases beyond the near-horizon region. This is because we can obtain the full expression of the $\lambda_{tc}$ and $\lambda_{nc}$ at any location of the radial coordinate. From \meqref{ltc}, we can write down the Lyapunov exponent $\lambda_{tc}$ as
\be
\begin{aligned}
\lambda_{tc}^2=&\frac{M^2-Q^2}{(\pi_\phi^2+r^2)^2}+\frac{\pi_\phi^4(3rQ^2(2M-r)+r^2M(2r-3M)-2Q^4)}{r^6(\pi_\phi^2+r^2)^2}
\\
&-\frac{\pi_\phi^2(3Q^4+2rQ^2(3r-5M)+r^2(6M^2-6Mr+r^2))}{r^4(\pi_\phi^2+r^2)^2}.
\end{aligned}\label{ltcrn}
\ee
From \meqref{lnc}, in terms of the RN black hole metric, $\lambda_{nc}$ can be rewritten as 
\be
\lambda_{nc}^2=\frac{2Q^4+r^2M(3M-2r)+3rQ^2(r-2M)}{r^6}.\label{lncrn}
\ee

\begin{figure}[htbp]
  \begin{subfigure}{2in}
    \centering
      \includegraphics[scale=0.25]{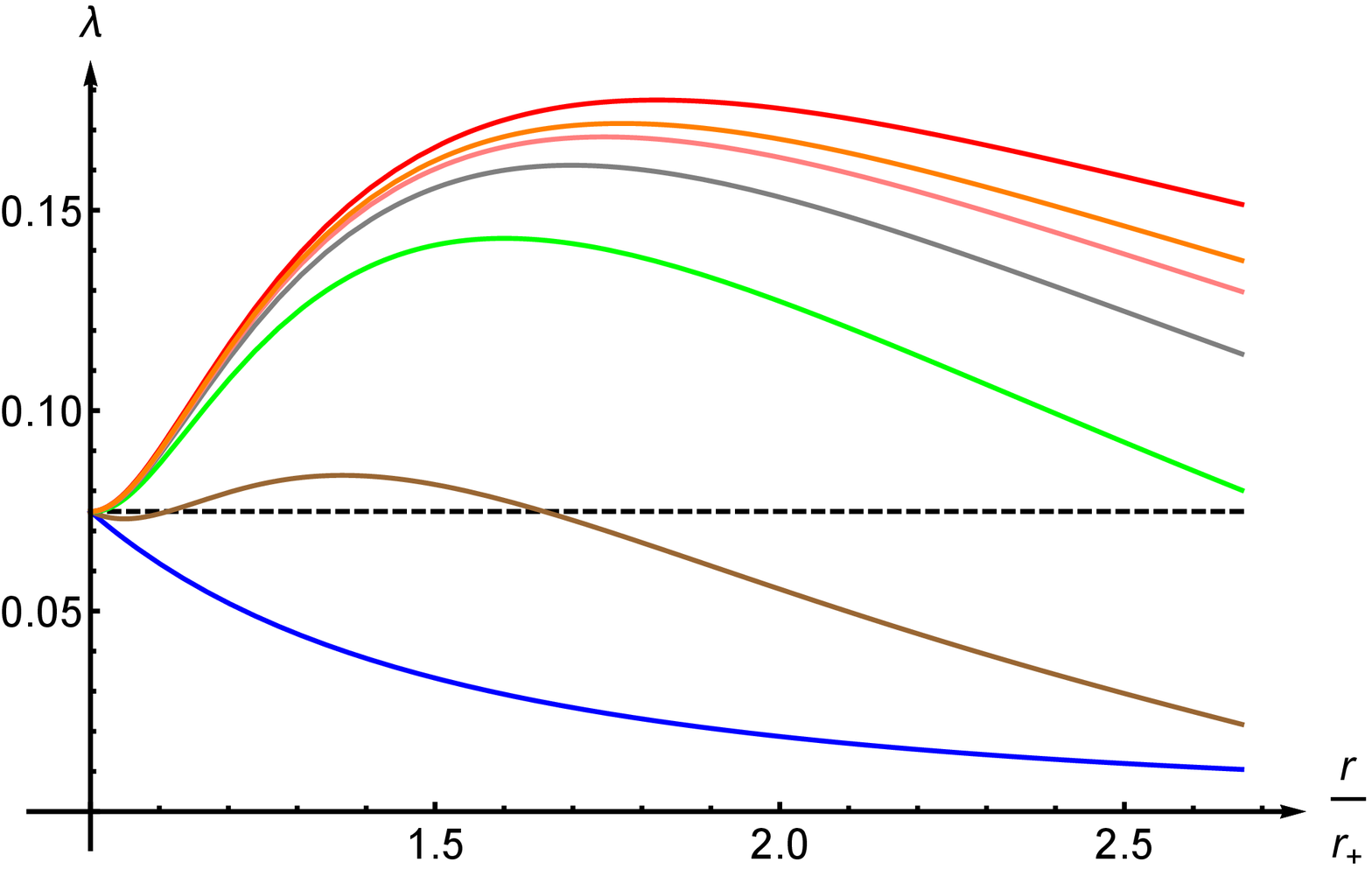}
      \caption{M/Q=1.004}\label{fig2a}
  \end{subfigure}
  \quad 
  \begin{subfigure}{2in}
    \centering
      \includegraphics[scale=0.25]{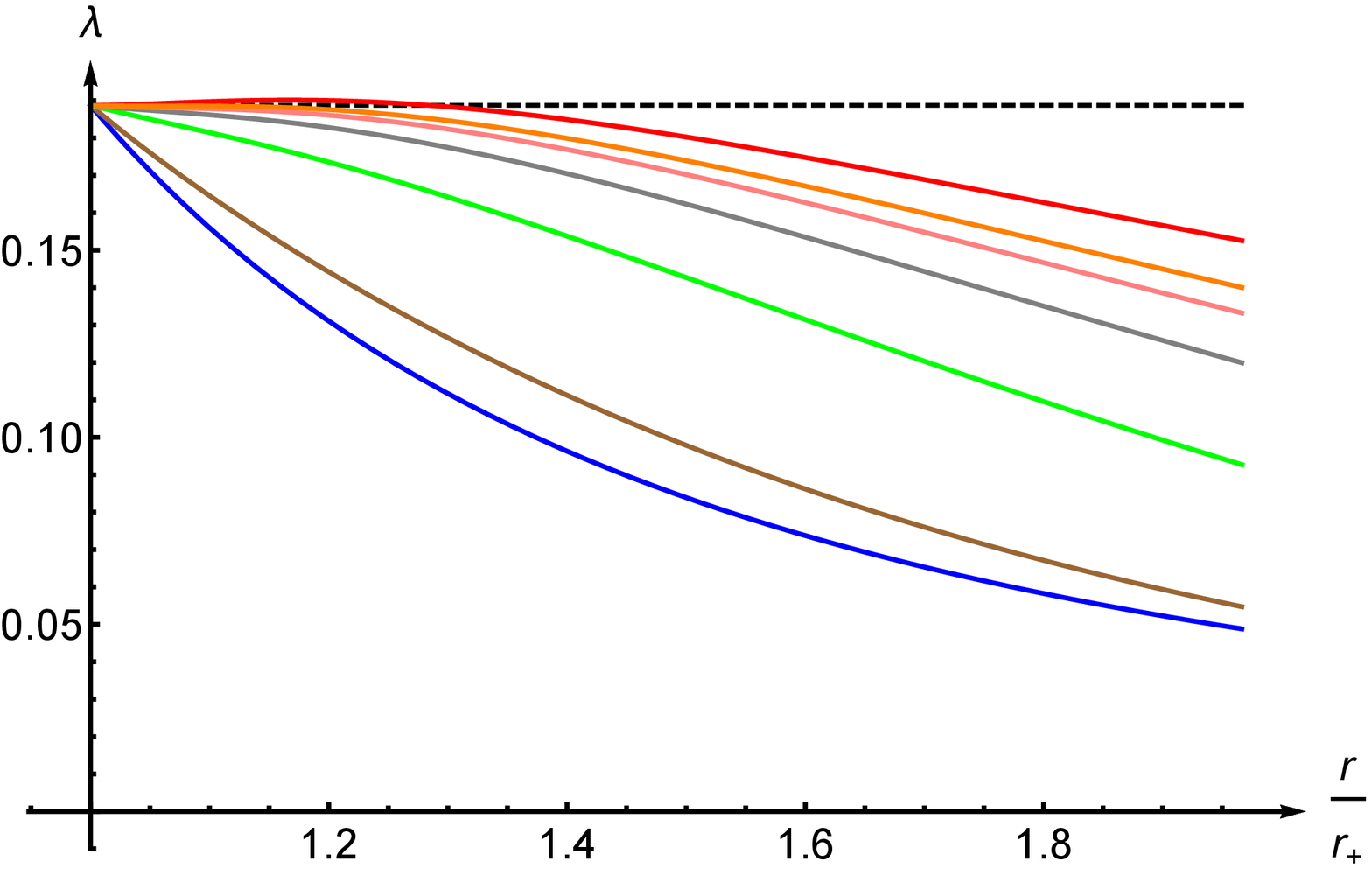}
      \caption{M/Q=1.1}\label{fig2b}
  \end{subfigure}
  \quad
  \begin{subfigure}{2in}
    \centering
      \includegraphics[scale=0.25]{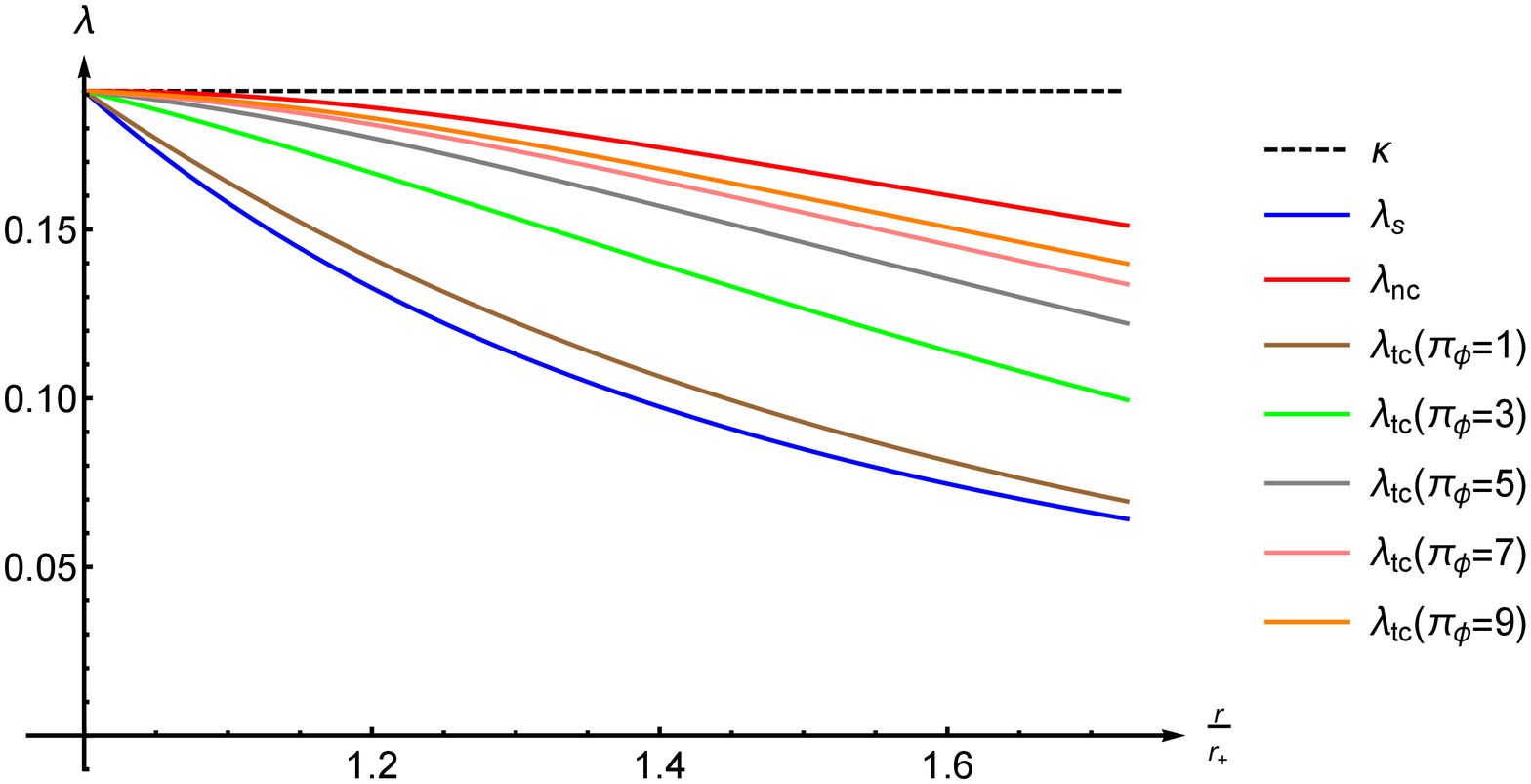}
      \caption{M/Q=1.2}\label{fig2c}
  \end{subfigure}
  \captionsetup{justification=raggedright}
  \caption{The Lyapunov exponents $\lambda_{tc}$, $\lambda_{nc}$ in \meqref{ltcrn} and \meqref{lncrn} as a function of $\frac{r}{r_+}$ for different mass-charge ratio of RN black holes, where the value of r range between horizon $r_+$ and the radius of the photon sphere $r_{ps}$. These can be used to support the near-horizon result of $\lambda$.}\label{fig2}   
\end{figure}

With \meqref{ltcrn} and \meqref{lncrn}, we draw the relationship between the Lyapunov exponent and the radius $r$ in \mpref{fig2}. As a control, we also draw the Lyapunov exponent of charged particles' static equilibrium\footnote{The expression is $\lambda_s^2=f^{\frac{3}{2}}\left((\sqrt{f})^{'}\frac{A_t^{''}}{A_t^{'}}-(\sqrt{f})^{''} \right)$} with blue line. We mainly focus on regions close to the horizon, so the radial range we study is from the outer horizon $r_+$ to the radius $r_{ps}$ of the photon sphere, where $r_{ps}=\frac{3M}{2}\left(1+\sqrt{1-\frac{8Q^2}{9M^2}} \right)$. In \mpref{fig2}, we show $\lambda_s$(static equilibrium), $\lambda_{nc}$(null circular motion), $\lambda_{tc}$(time-like circular motion) corresponding to different angular momentum $L=1,3,5,7,9$, and a horizontal line $\kappa$ which is used to test whether the chaos bound is violated. For all time-like circular trajectories in \mpref{fig2}, a common feature can be found that the Lyapunov exponent of time-like motion increases as the angular momentum goes up, which indicates that angular momentum will increase the instability of time-like circular motion near the horizon.

In \mpref{fig2a}, for the time-like circular orbits with angular momentum $L=1,3,5,7,9$, $\lambda_{tc}$ are greater than $\kappa$ and $\lambda_{nc}$ also exceeds $\kappa$, which means the chaos bound is violated. From \mpref{fig2b}, the violation of the chaos bound exists in the Lyapunov exponent $\lambda_{nc}$ of null circular motion, but not for time-like circular motion, because the angular momentum is not large enough. The effect of angular momentum on $\lambda_{tc}$ is not as significant as that caused by the changing in $M/Q$. The ratio $M/Q$ has a sharp influence on the behavior of Lyapunov exponents. In \mpref{fig2c}, $\lambda_{tc}$ and $\lambda_{nc}$ are smaller than $\kappa$, and the chaos bound is not violated. These results verify our above interpretations about the violation of chaos bound. Also, whether it is static equilibrium, time-like circular motion or null circular motion, their Lyapunov exponents always have $\lambda=\kappa$ on the horizon, which means the chaos bound is saturated. However, outside the horizon, as the angular momentum increases from 0 to infinity, the Lyapunov exponent $\lambda_{tc}$ of the time-like circular motion changes from $\lambda_{s}$ of the static equilibrium to $\lambda_{nc}$ of the null circular motion, and the larger the angular momentum of the time-like circular motion, the larger the Lyapunov exponent.

After all, the violation of chaos bound exists in the circular motion of charged particles near the RN black hole. In our calculation, this phenomenon appears in the RN black holes with $1<M/Q<1.1547$, which should relate to the property of black holes. By comparing the examples of $M/Q=1.004$ and $M/Q=1.1$, it can be clearly seen that the smaller the values of $M/Q$, the more likely chaos bound can be violated. In other words, we can infer that the closer to the extremal RN black hole, the more obvious the chaos bound can be violated, and this may have rich physical significance worthy of further study.

\section{Reissner-Nordstr\"om Anti-de Sitter (RN-AdS) black hole}\label{RNAdS}
In this section, we consider a toy model of the circular motion of charged particle near 4-dimensional Reissner-Nordstr\"om Anti-de Sitter (RN-AdS) black hole which has a negative cosmological constant $\Lambda_0=-3/\ell^2$. $\ell$ is the radius of AdS spacetime. The metric function $f(r)$ and the electric potential function $A_t(r)$ of the RN-AdS black hole are given by
\be
f(r)=\frac{r^2}{\ell^2}+k-\frac{2 M}{r}+\frac{Q^2}{r^2},\qquad A_t(r)=\frac{2Q}{r},
\ee
where $M$ and $Q$ are the mass and charge of black hole, $k$ is the topological parameter, which can take (-1, 0, 1). There are two black hole horizons $r_-$ and $r_+$ ($r_+ > r_-$),  the parameters ($M$, $Q$) and the surface gravity $\kappa$ can be expressed as
\be
\begin{aligned}
M=&\frac{(r_-+r_+)(k\ell^2+r_-^2+r_+^2)}{2\ell^2}, 
\\
Q=&\frac{\sqrt{r_-r_+(k\ell^2+r_-^2+r_-r_++r_+^2)}}{\ell},
\\
\kappa=&\frac{(r_+-r_-)(k\ell^2+r_-^2+2r_-r_++3r_+^2)}{2\ell^2r_+^2}.
\end{aligned}
\ee
To simplify subsequent calculations, we consider the specific values of the parameters ($k$ and $\ell$):
\be
k=1,\qquad \ell=1.
\ee
By taking $Q = 1$, it indicates that for a non-extremal RN-AdS black hole, its mass $M$ should satisfy $M>1.23132$. 

\subsection{Near-horizon analysis}

The near-horizon expansion parameters ($\gamma_{t1}$, $\gamma_{t2}$, $\gamma_{n1}$, $\gamma_{n2}$) can be obtained as
\be
\begin{aligned}
\gamma_{t1}=&-\frac{4(Mr_+-Q^2+r_+^4)^2}{r_+^5(\pi_\phi^2+r_+^2)},
\\
\gamma_{t2}=&\frac{2(Q^2-r_+(M+r_+^3))(\pi_\phi^4(3Mr_+-6Q^2)+r_+^4(5Q^2-5Mr_++r_+^4)+\pi_\phi^2r_+^2(7r_+^4+4Mr_+-7Q^2))}{r_+^8(\pi_\phi^2+r_+^2)^2},
\\
\gamma_{n1}=&0,
\\
\gamma_{n2}=&-\frac{6(Mr_+-2Q^2)(Mr_+-Q^2+r_+^4)}{r_+^8}.
\end{aligned}\label{grnads}
\ee

With $\gamma_{n2}$ in \meqref{grnads}, the condition $M<1.75225$ equals to $\gamma_{n2}>0$ which means that the chaos bound can be violated by the null circular motion of charged particle outside the RN-AdS black hole. Therefore, in this section we consider taking $M$ with the values (1.3, 1.6, 1.8) as $Q=1$. For the RN-AdS black hole with $M=1.3,\ 1.6$, the null circular motion will exceed the chaos bound, while it does not in the case with $M=1.8$. 
In \mtref{gnads}, we show the parameters $\gamma_{n1}$ and $\gamma_{n2}$ corresponding to the null circular motion outside different RN-AdS black holes. $\gamma_{n1}$ in all cases are 0, and for the RN-AdS black holes with $M=1.3, \ 1.6$, we can find $\gamma_{n2}>0$ which indicates the chaos bound can be violated by null circular motion outside the black holes. For the case of $M=1.8$, no violation exists because of $\gamma_{n2}<0$. 

\begin{table}[htbp]
    \centering
    \begin{tabular}{ccc} 
        \hline
        RN-AdS black holes&$\gamma_{n1}$&$\gamma_{n2}$\\
         \hline
         M=1.3 &0  &13.83918 \\ 
         \hline
         M=1.6 &0  &2.20714\\
         \hline
         M=1.8 &0 &-0.49644 \\ 
         \hline
    \end{tabular}
    \captionsetup{justification=raggedright}
    \caption{The parameters ($\gamma_{t1}$ and $\gamma_{t2}$) of null circular motion near different RN-AdS black holes where $Q=1$.}\label{gnads}
\end{table}

\begin{figure}[htbp]
  \begin{subfigure}{3in}
    \centering
      \includegraphics[scale=0.38]{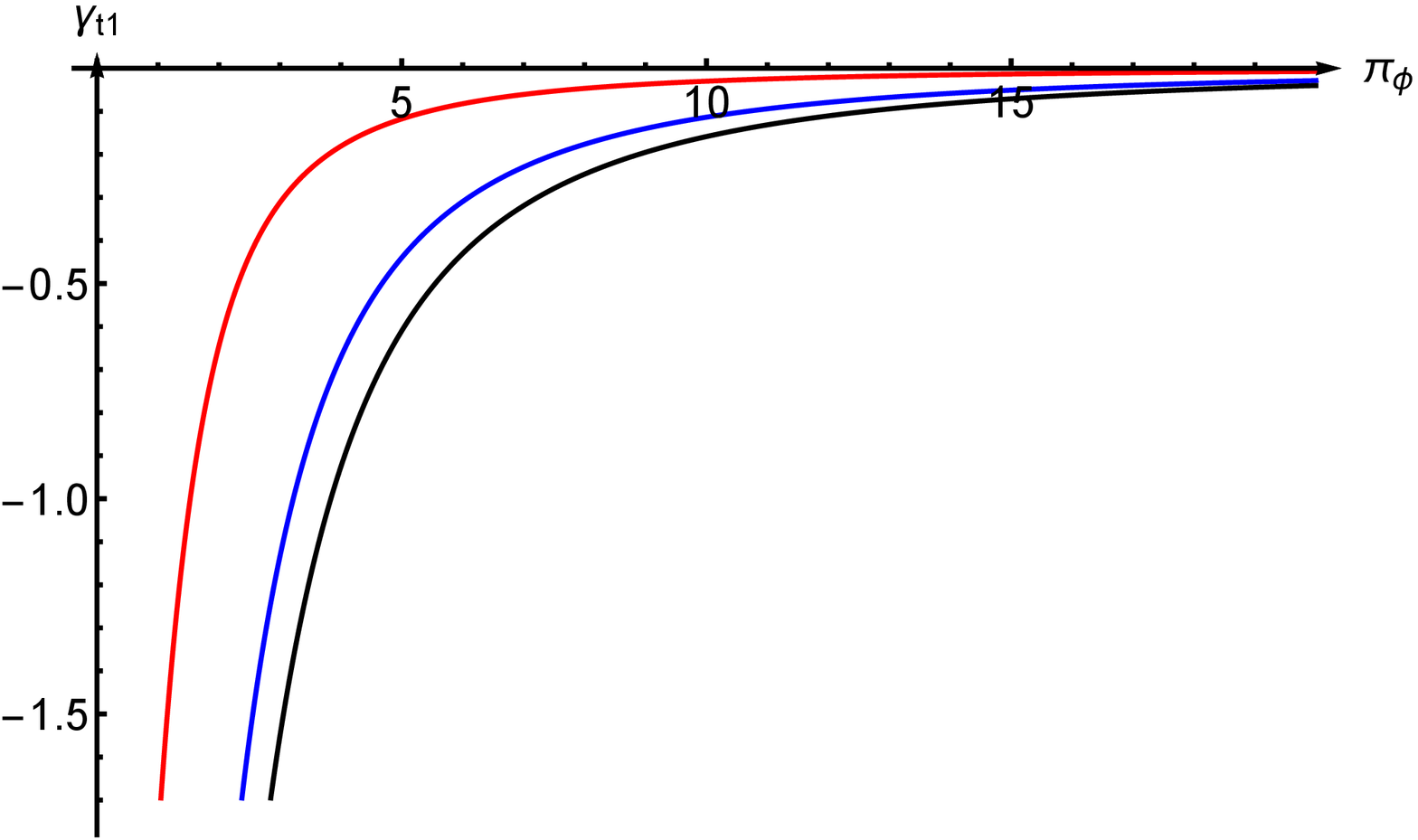}
      \caption{$\gamma_{t1}$ as a function of the angular momentum $\pi_\phi$.}\label{fig3a}
  \end{subfigure}
  \qquad 
  \begin{subfigure}{3in}
    \centering
      \includegraphics[scale=0.37]{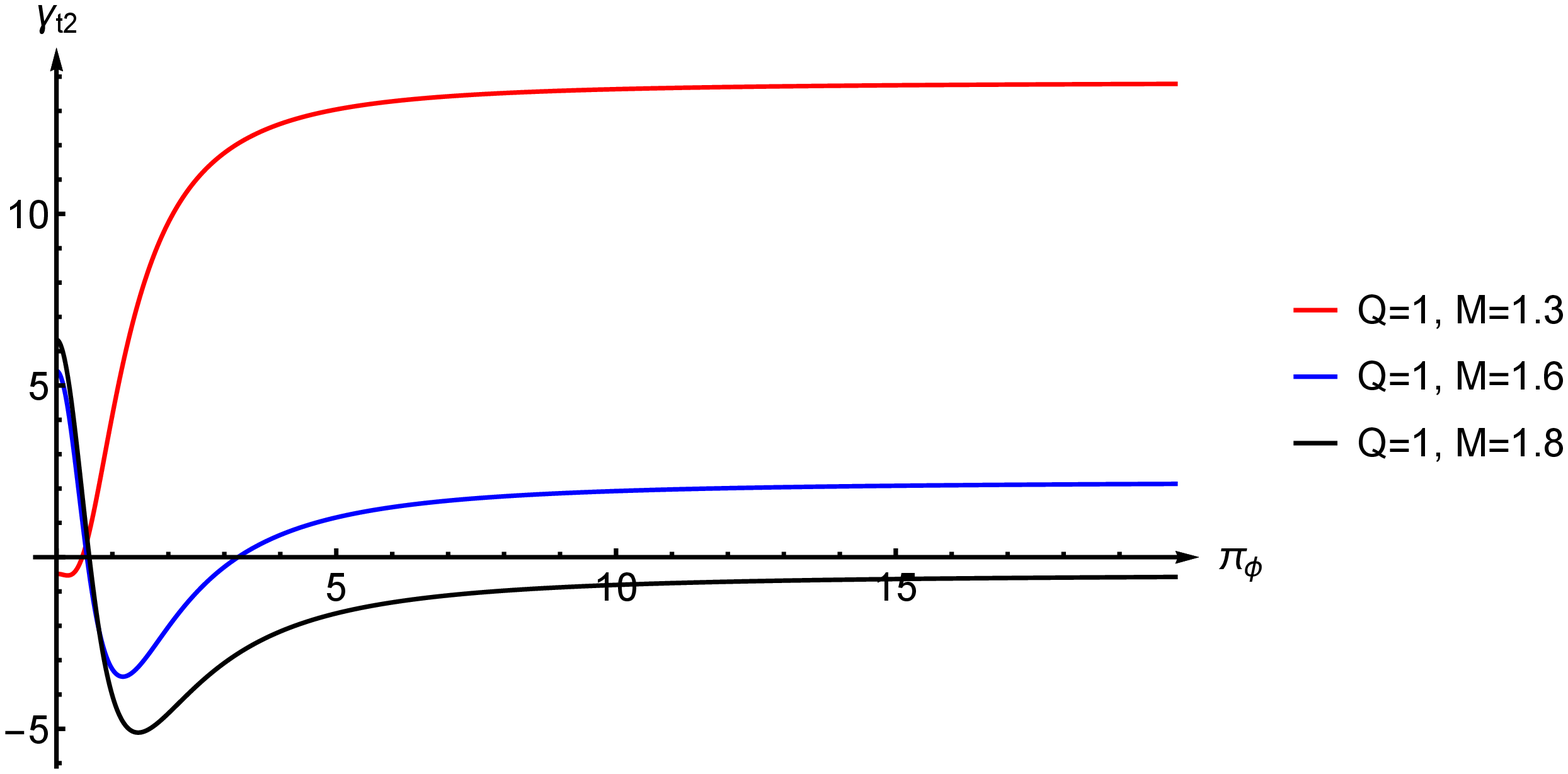}
      \caption{$\gamma_{t2}$ as a function of the angular momentum $\pi_\phi$.}\label{fig3b}
  \end{subfigure}
  \captionsetup{justification=raggedright}
  \caption{The parameters ($\gamma_{t1}$, $\gamma_{t2}$) as a function of the angular momentum $\pi_\phi$ for different mass $M$ and charge $Q$ of RN-AdS black holes.}\label{fig3}   
\end{figure}

We show the relationship between parameters ($\gamma_{t1}$, $\gamma_{t2}$) and time-like circular motion's angular momentum $\pi_\phi$ in \mpref{fig3}. In \mpref{fig3a}, we can find that $\gamma_{t1}$ is negative for all cases, but approaching 0 as in the large $\pi_\phi$ limit. The smaller the $M/Q$ ratio, the faster $\gamma_{t1}$ approaches zero. On the other hand, as the angular momentum increases, $\gamma_{t2}$ converges to a constant value ($\gamma_{n2}$) in \mpref{fig3b}. These properties are the same as in the RN black hole in flat spacetime, so we can make the interpretation that near the RN-AdS black holes with $M=1.3,\ 1.6$, the time-like circular motion can exceed the chaos bound in the large $\pi_\phi$ limit, and it cannot in the background of RN-AdS black hole with $M=1.8$. The critical ratio $M/Q$ is $M/Q<1.75225$. Similarly, $\gamma_{t1}$ converges to 0 faster as the RN-AdS black hole gets closer to the extremes, implying that the violation of the chaos bound may occur at lower temperature. 
\subsection{Beyond the near-horizon region}

Next, we analyze the Lyapunov exponent of the circular motion by plotting the expressions of $\lambda_{tc}$ and $\lambda_{nc}$, where the near-horizon expansion is not considered. The Lyapunov exponents of charged particle circular motion can be obtained from the full expression \meqref{ltc} and \meqref{lnc} by substituting the metric of RN-AdS black hole 
\be
\begin{aligned}
\lambda^2_{tc}=&-\frac{\pi_\phi^2(3Q^4+2Q^2r_+(7r_+^3+3r_+-5M)+r_+^2(6M^2+r_+^2+6r_+^4+3r_+^6-6M(r_++r_+^3)))}{r_+^4(\pi_\phi^2+r_+^2)^2}
\\
&-\frac{\pi_\phi^4(2Q^4+3Q^2r_+(2r_+^3+r_+-M)+r_+^2(3M^2+r_+^4-2M(r_++3r_+^3)))}{r_+^6(\pi_\phi^2+r_+^2)^2}
\\
&+\frac{M^2+8Mr_+^3-3r_+^4-6Q^2r_+^2-Q^2-2r_+^6}{(\pi_\phi^2+r_+^2)^2},
\end{aligned}\label{rnadsltc}
\ee
\be
\lambda^2_{nc}=-\frac{\pi_\phi^4(2Q^4+3Q^2r_+(2r_+^3+r_+-2M)+r_+^2(3M^2+r_+^4-2Mr_+-6Mr_+^3))}{r_+^6(\pi_\phi^2+r_+^2)^2}.\label{rnadslnc}
\ee

\begin{figure}[htbp]
    \begin{subfigure}{2in}
      \centering
        \includegraphics[scale=0.25]{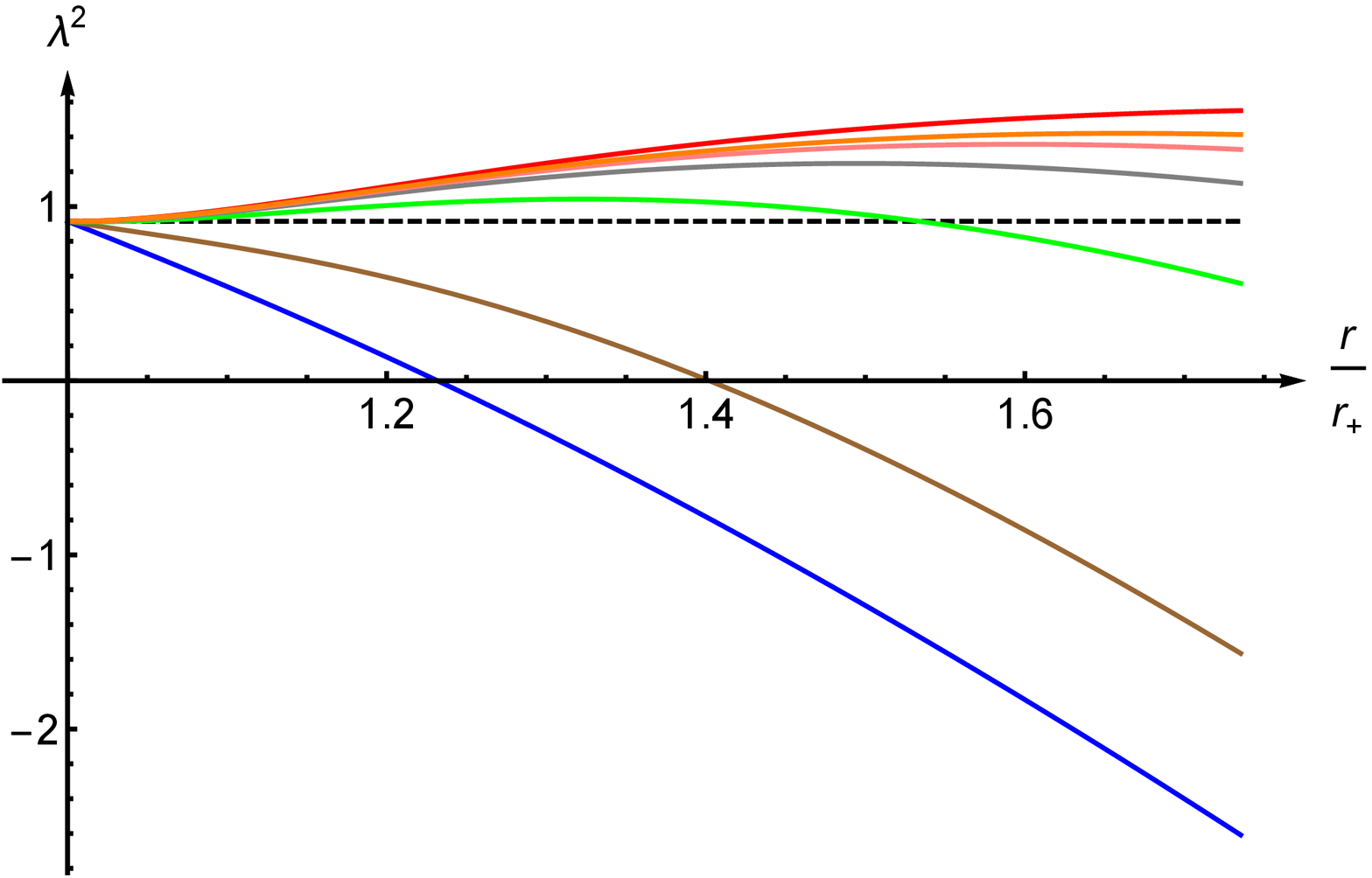}
        \caption{Q=1, M=1.3}\label{fig4a}
    \end{subfigure}
    \quad 
    \begin{subfigure}{2in}
      \centering
        \includegraphics[scale=0.25]{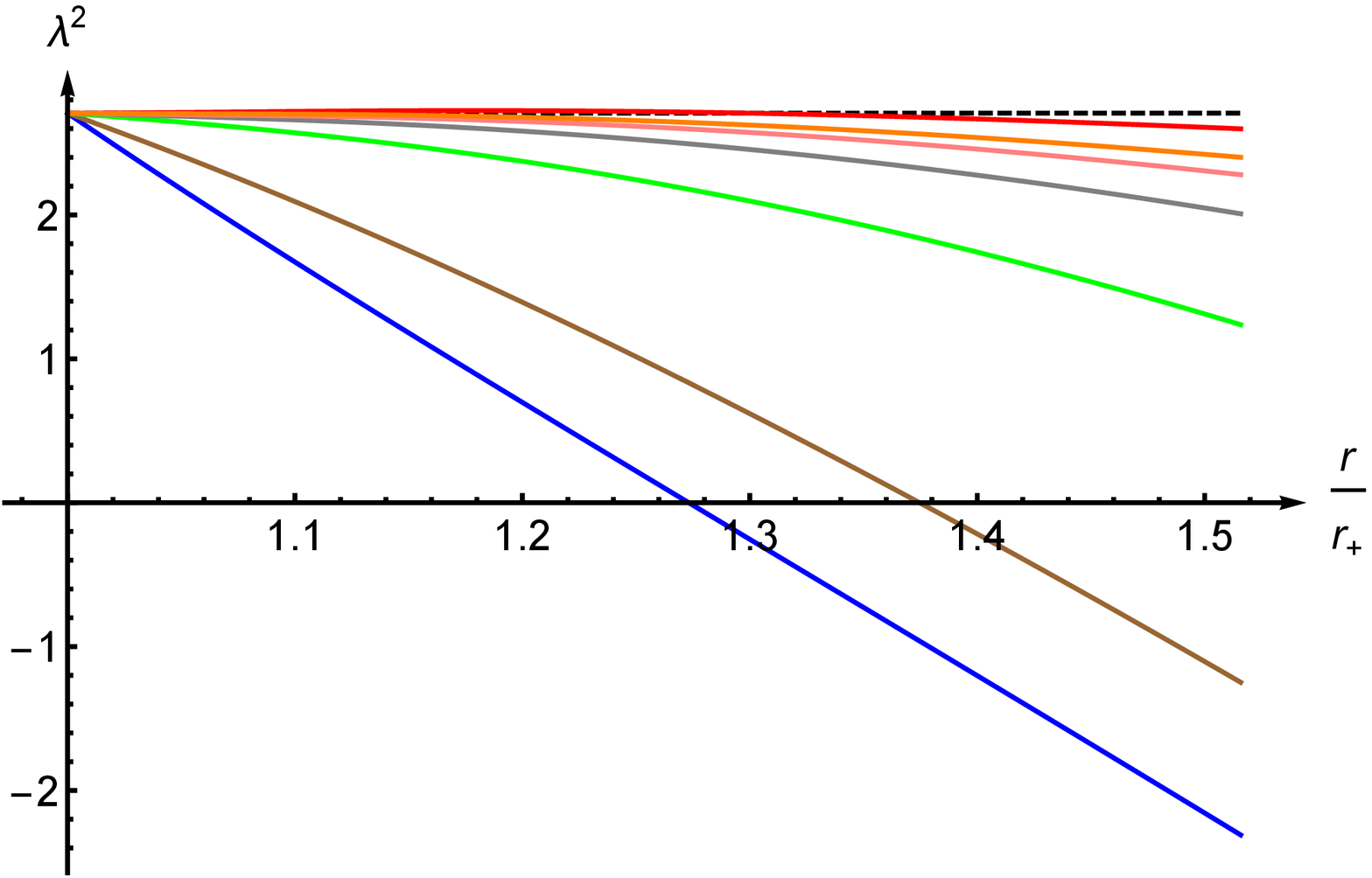}
        \caption{Q=1, M=1.6}\label{fig4b}
    \end{subfigure}
    \quad
    \begin{subfigure}{2in}
      \centering
        \includegraphics[scale=0.25]{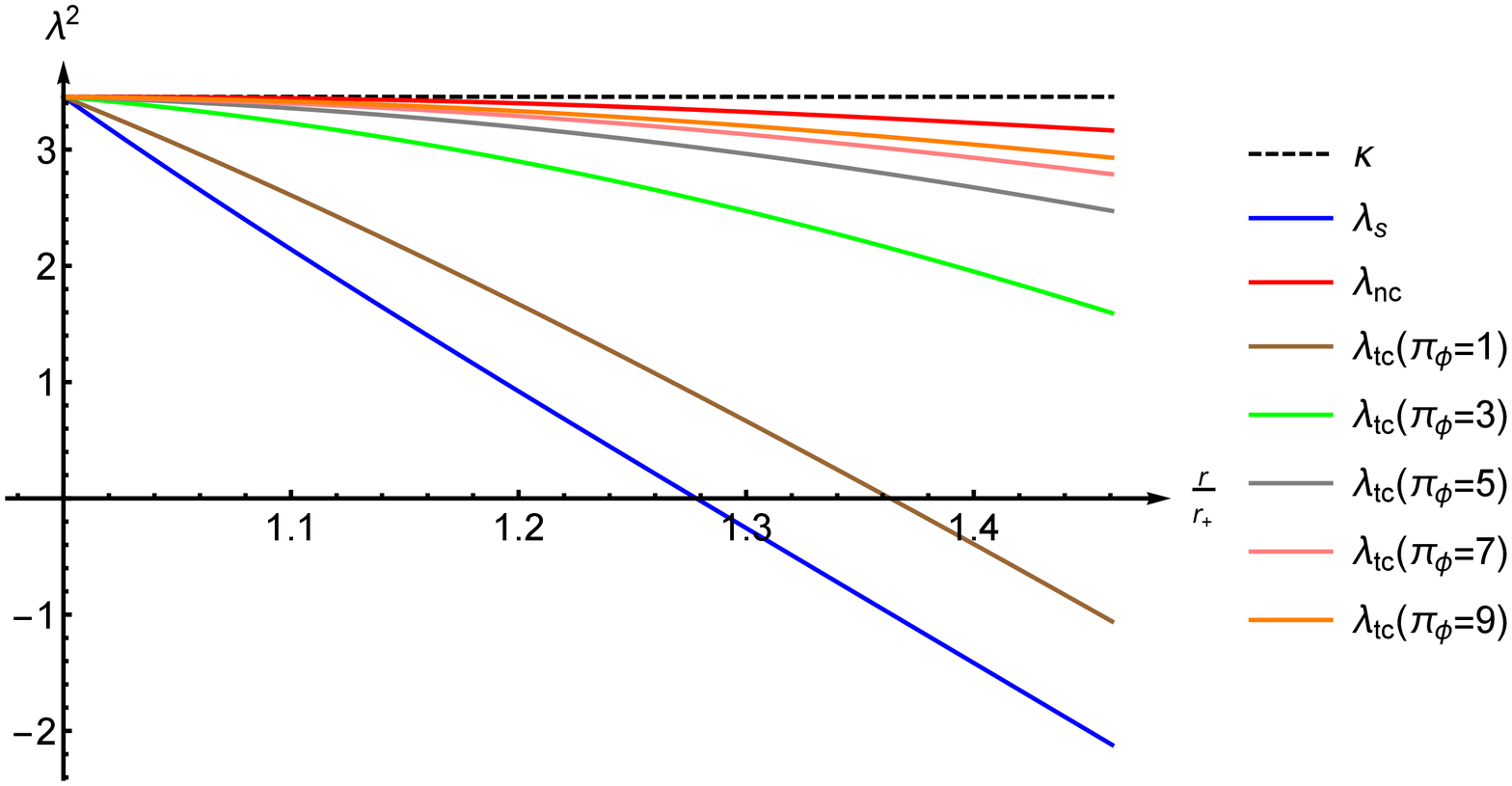}
        \caption{Q=1, M=1.8}\label{fig4c}
    \end{subfigure}
    \captionsetup{justification=raggedright}
    \caption{The Lyapunov exponents $\lambda_{tc}$, $\lambda_{nc}$ in \meqref{rnadsltc} and \meqref{rnadslnc} as a function of $\frac{r}{r_+}$ for different mass-charge ratio of RN-AdS black holes. The value of r is considered between the horizon $r_+$ and the radius of the photon sphere $r_{ps}$.}\label{fig4}   
\end{figure}

Then we discuss the Lyapunov exponent outside different RN-AdS black holes as the function of the radial coordinater $r$. We draw the Lyapunov exponent near different RN-AdS black holes in \mpref{fig4}. In \mpref{fig4a}, the null circular motion and time-like circular motion with large angular momentum can violate the chaos bound. In \mpref{fig4b}, the chaos bound is exceeded by the null circular motion, and not for the time-like circular motion because their angular momentum is not large enough. For the RN-AdS black hole with ($Q=1,\ M=1.8$) in \mpref{fig4c}, there is no violation of the chaos bound, just as we can know from \mpref{fig3}. For the RN-AdS black holes, the Lyapunov exponent of charged particles' static equbilium $\lambda_s$ and time-like circular motion $\lambda_{tc}$ can be imaginary, which is different from in RN black holes. This may be the effect of the AdS background.

Our calculation for the RN-AdS yields the same conclusions as in the RN black hole. For RN-AdS black hole, the angular momentum increases the instability of the time-like circular motion near the horizon. The possibility and conditions for time-like circular motion with large angular momentum and null circular motion exceeding the chaos bound can also be described by the parameters $(\gamma_{t1},\ \gamma_{t2},\ \gamma_{n1},\ \gamma_{n2})$. These conclusions are universal and not affected by the AdS background.

\section{Conclusion}\label{conclusion}
In summary, the circular motion of the massive and massless charged particles has been studied. The Lyapunov exponent of the circular motion is calculated by using the Jacobian matrix to explore the validation of the chaos bound near black holes. From the general calculation, we can see that no matter the massive particles or massless particles, their circular motion at the horizon can saturate the chaos bound. We also define some parameters to discuss the relationship between the Lyapunov exponent $\lambda$ and the surface gravity $\kappa$ (chaos bound). Under the conditions that the mass-charge ratio of RN black hole meet $1<M/Q<1.1547$, the chaos bound can be violated locally in the circular motion of charged particles (for RN-AdS black hole, it is $1.23132<M/Q<1.75225$). It is the first time to find that the chaos bound $\lambda \leq \kappa$ can be violated locally for circular motions of RN (AdS) black holes.

In particular, the RN black holes with different $M/Q$ have been investigated. The calculations have shown some interesting conclusions. Compared with the static equilibrium of charged particles, the angular momentum of circular motion will lead to the increase of the Lyapunov exponent in particles' radial equilibrium. The larger the angular momentum, the larger the Lyapunov exponent. When the angular momentum tends to infinity, it corresponds to the null circular motion. Meanwhile, the results indicate that the speed of light provides a maximum value of the Lyapunov exponent. In addition, we also find that the circular motion of charged particles near the horizon can violate the chaos bound. Our calculations show that near the RN black hole with $1<M/Q<1.1547$, the chaos bound can be violated by the null circular motion and the time-like circular motion with large angular momentum. Then RN-AdS black holes are considered to verify the universality of these conclusions. The instability of time-like circular motion increases with the growth of angular momentum. As $Q=1$, the time-like circular motion with large angular momentum and null circular motion near the RN-AdS black holes which satisfy $1.23132 < M < 1.75225$ can exceed the chaos bound.

It is worth noting that for the RN black holes close to the extremal RN black hole, the violation of chaos bound will be more obvious. Due to the temperature dependence of the chaos bound, its violation in the near-extremal RN black hole should be determined by some properties of the black hole, and this unknown connection is undoubtedly worthy of our further study. Actually, the study on the dynamical stability of the extremal RN black hole \cite{Hod:2012wmy,Myung:2018vug} may help us. The stability studies of some cases of extremal charged black holes \cite{Hod:2018fet,Ge:2008ni,Ge:2009eh} may help us understand the exceeding of chaos bound by the circular motion of charged particles outside the horizon. Many black hole thermodynamic problems may be relevant, such as the holographic heat engine \cite{Sun:2021yen,Wei:2017vqs,Fang:2017nse,Cai:2004pz}. Our calculations in this paper do not consider the backreaction of test particles on the black hole background. For the unstable extremal RN black hole and the test particles with large angular momentum, it may be more meaningful to discuss the backreaction of the perturbation on the black hole background spacetime. Related discussions may lead to more interesting results.

What we need to emphasize is that this violation of chaos bound under the background of the near-extremal RN black hole is not reflected in the static equilibrium of charged particles, and our calculations on circular motion highlight this point. The null circular motion near the horizon is very likely to help us further explore the connection between the chaos bound and the quasi-normal modes, pole-skipping and shock waves. After all, the study of the circular motion of charged particles outside a charged black hole helps us examine the conclusion that the radial equilibrium of particles saturates the chaos bound at the horizon again. At the same time, the violation of the chaos bound in the near-extremal RN black hole indicates deep connections among the chaos bound, black hole thermodynamic and other issues. 

\section*{Acknowledgement} We would like to thank Long Cheng and YiLi Wang for helpful discussions. The work was partially supported by NSFC, China (grant No.11875184).\\

\section*{Note Added}
When we finished the main part of this paper, we noticed the preprint \cite{Kan:2021blg}, where the authors discuss the bound of the Lyapunov exponent in charged particle motion. That is an interesting work on understanding the chaos bound, and our calculations discuss similar situations. But our work is done independently. The difference is we use the Jacobian matrix method to calculate the Lyapunov exponent instead of the effective potential analysis. We pay attention to the circular motion of charged particles near charged black holes and define some parameters under the near-horizon expansion to predict whether the chaos bound can be violated by the circular motion.

\section*{Appendix}
\begin{appendix}
\section{More discussion on $\gamma_{t1}$}
Discussing in the background of RN black holes and RN-AdS black holes, we find that in the background closer to the extreme black hole, the $\gamma_{t1}$ of the time-like circular motion tends to 0 faster with the growth of angular momentum, which means that the chaos bound is more likely to be violated by the time-like circular motion. In this section, we examine more cases to show that $\gamma_{t1}$ of the time-like circular motion tends to 0 faster the closer to the extreme black hole.

\begin{figure}[htbp]
  \centering
        \includegraphics[scale=0.75]{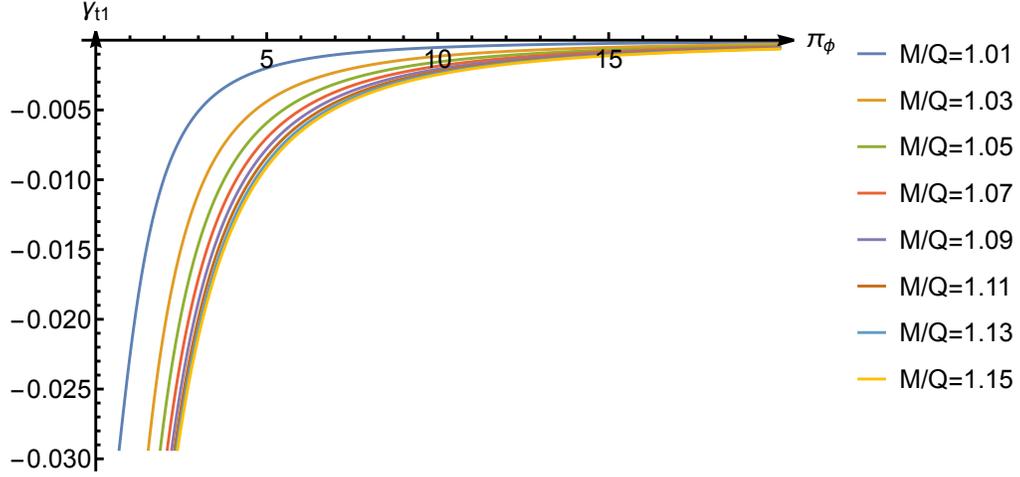}
    \captionsetup{justification=raggedright}
    \caption{The parameters $\gamma_{t1}$ as a function of the angular momentum $\pi_\phi$ for different mass-charge ratio $M/Q$ of RN black holes, when $Q=1$.}\label{fig5}   
\end{figure}

\begin{figure}[htbp]
  \centering
        \includegraphics[scale=0.75]{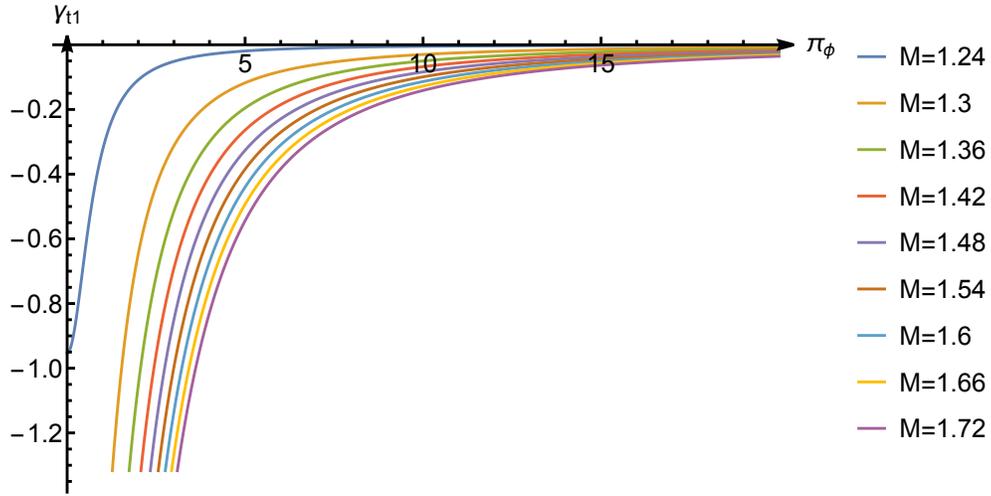}
        \captionsetup{justification=raggedright}
    \caption{The parameters $\gamma_{t1}$ as a function of the angular momentum $\pi_\phi$ for different RN-AdS black holes, when $Q=1$.}\label{fig6}   
\end{figure}

In \mpref{fig5}, we discuss the relationship between $\gamma_{t1}$ of the time-like circular motion and the angular momentum $\pi_\phi$ under the background of RN black holes. Different mass-charge ratio $M/Q$ are considered, the closer $M/Q$ is to 1 (extreme RN black hole), the faster $\gamma_{t1}$ tends to 0. $\gamma_{t1}$ of the time-like circular motion near RN-AdS black holes is shown in \mpref{fig6}. As $Q=1$, the extremal RN-AdS black hole satisfies $M=1.23132$. So, we can see that $\gamma_{t1}$ of the time-like circular motion will tend to 0 more faster as the RN-AdS black hole approaches the extreme black hole.

As shown in \mpref{fig5} and \mpref{fig6}, $\gamma_{t1}$ of the time-like circular motion near RN black holes and RN-AdS black holes tends to 0 faster when approaching the extreme black hole, this implies that the chaos bound is violated more easily. This behavior implies an intrinsic relationship between black hole chaos and problems in black hole thermodynamics, dynamics, etc.
\\
\\
\\
\\

\end{appendix}

\end{document}